\documentclass[12pt]{article}
\usepackage{color,epsfig,amsmath,amssymb,graphicx}
\definecolor{violet}{rgb}{0.4,0,0.2}
\definecolor{vert}{rgb}{0,0.6,0.0}
\definecolor{navy}{rgb}{0.0,0.0,0.6}
\definecolor{orange}{rgb}{0.6,0.4,0.0}
\definecolor{bleu}{rgb}{0.3,0.0,0.8}
\definecolor{spin}{rgb}{0.3,0.6,0.0}

\def\bb{\large\color{black}  $ }  \def\fb{ $  }
\def\be{\large\begin{equation}\color{black} }
\def\fe{\end{equation}}
\def\rf{\color{black} (\ref }
\def\fr{)\,\color{navy}\, }

\def\pp{{\color{orange}p}}
\def\Deltav{{\color{violet}{\mit\Delta}}}
\def\vv{{\color{violet}v}}

\def\calA{{\color{violet}{\cal A}}}
\def\calK{{\color{violet}{\cal K}}}

\def\ss{{\color{vert}s}}
\def\nn{{\color{vert}n}}

\def\hnn{{\color{violet}\hat n}}
\def\kk{{\color{vert}k}}

\def\Lamb{{\color{red}\Lambda}}
\def\Pssi{{\color{red}\Psi}}
\def\rrho{{\color{red}\rho}}
\def\aalpha{{\color{red}\alpha}}
\def\mm{{\color{red}m}}

\def\mmu{{\color{red}\mu}}
\def\cchi{{\color{red}\chi}}
\def\ppi{{\color{red}\pi}}
\def\hba{{\color{red}\hbar}}

\def\UU{{\color{red}U}}

\def\pp{{\color{red}p}}

\def\uu{{\color{violet}u}}

\def\TT{{\color{red}T}}

\def\hHH{{\color{red}{\hat H}}}

\def\Thet{ {\color{red} \Theta}}
\def\ZZ{ {\color{orange} Z}}
\def\AA{ {\color{black} A}}
\def\Alf{ {\color{black} a}}
\def\rmn{ {\rm n}} \def\rmp{{\color{orange}\rm p}}
\def\rmf{ {\rm f}} \def\rmb{ {\rm b}} \def\rmc{ {\rm c}}
\def\rmI{ {\rm I}} \def\rmfc{ {\rm fc}}

\def\spose#1{\hbox to 0pt{#1\hss}}\def\lta{\mathrel{\spose{\lower 3pt\hbox
{$\mathchar"218$}}\raise 2.0pt\hbox{$\mathchar"13C$}}}  \def\gta{\mathrel
{\spose{\lower 3pt\hbox{$\mathchar"218$}}\raise 2.0pt\hbox{$\mathchar"13E$}}}

\begin{document}

\title{\color{navy}\bf Entrainment coefficient and effective mass for
conduction neutrons in neutron star crust:  Macroscopic treatment.}

\author { {\bf
Brandon Carter$^{a}$,
Nicolas Chamel$^{a,b}$,
Pawel Haensel$^{a,b}$ }
\\ \hskip 1 cm\\   \\
$$ $^{a}$ Observatoire de Paris, 92195 Meudon, France \\
(Brandon.Carter@obspm.fr, Nicolas.Chamel@obspm.fr)$$
\\ $$ $^{b}$ N. Copernicus Astronomical Center, Warsaw, Poland \\
 (haensel@camk.edu.pl)$$ }

\date{\it 28 January, 2006}

\maketitle

{\catcode `\@=11 \global\let\AddToReset=\@addtoreset}
\AddToReset{equation}{section}
\renewcommand{\theequation}{\thesection.\arabic{equation}}

\color{navy}

\bigskip 
{\bf Abstract.}

Phenomena such as pulsar frequency glitches are believed to be attributable
to differential rotation of a current of ``free'' superfluid neutrons at 
densities above the ``drip'' threshold in the ionic crust of a neutron star. 
Such relative flow is shown to be locally describable by adaption of a
canonical two fluid treatment that emphasizes the role of the momentum 
covectors constructed by differentiation of action with respect to the
currents, with allowance for stratification whereby the ionic number current 
may be conserved even when the ionic charge  number {\bb \ZZ\fb} is altered 
by beta processes. It is demonstrated that the gauge freedom to make 
different choices of the chemical basis determining which neutrons are
counted as ``free'' does not affect their ``superfluid'' momentum covector, 
which must locally have the form of a gradient (though it does affect the 
``normal'' momentum covector characterising the protons and those neutrons 
that are considered to be ``confined'' in the nuclei). It is shown how the 
effect of ``entrainment'' (whereby the momentum directions deviate from
those of the currents) is controlled by the (gauge independent) mobility 
coefficient {\bb\calK\fb}, estimated in recent microscopical quantum
mechanical investigations, which suggest that the corresponding (gauge 
dependent) ``effective mass'' {\bb\mm_\star\fb} of the free neutrons can 
become very large in some layers. The relation between this treatment of 
the crust layers and related work (using different definitions of 
``effective mass'') intended for the deeper core layers is discussed.

\section{Introduction}

As a prerequisite for the quantitative analysis of the role of differential
rotation in the angular momentum transfer mechanism that is 
thought~\cite{Ruderman} to be responsible for phenomena such as pulsar 
glitches, the purpose of this article is to describe the adaptation to the 
special circumstances pertaining to the inner crust of a neutron star of the 
kind of non relativistic two constituent fluid formalism that has already 
been applied~\cite{Lindblom00, PrixComAnd} to the global description of
the main bulk of the star, in which the basic constituents are relatively 
moving currents of protons and neutrons, using a treatment of the
non-dissipative kind~\cite{Mendell91,Prix01} that has recently been developed
in a variational formulation \cite{Rieutord02, Prix02, AndComGro} that, 
although non-relativistic, can nevertheless
be very conveniently expressed in a fully covariant manner~\cite{CC03,CC04}.
(A non-dissipative treatment is useful as a good
first approximation in view of the superfluidity of the neutrons, but in 
any case, after the appropriate action function has been obtained in the 
manner described below, it will be possible to use it as an equation of 
state in a less idealised model of the kind~\cite{CC05} needed to allow for 
dissipation and for the relativistic effects~\cite{LanSedCar} that are 
important at a global level.)

Such a two fluid treatment does not take account of the third independent 
current, namely that of the electrons, that is of dominant importance for 
electromagnetic effects, but which makes a relatively negligible
contribution to the Newtonian mass transport effects under consideration 
here. Allowance will however be made here for an important special feature 
of the crust layers (providing stratification that contributes to stability 
against convection) namely the clustering of the protons into ions, whose 
number current can remain conserved even after allowance for possible 
variation of the ionic charge number {\bb\ZZ\fb} due to weak interactions
whereby protons are transformed into neutron or vice versa. Another special 
stabilising feature of the crust that will also need to be allowed for, but
that will be left for subsequent work is that of the (relatively small) 
stress anisotropy that can arise from the elastic solidity \cite{CC06} of 
the crust, or from strong magnetic fields~\cite{CCC05}. 

The present article is particularly concerned rather with allowance for the 
effect of relative ``entrainment'' \cite{Andreev76} between on one hand the 
effectively free neutrons that will be present above the ``neutron drip'' 
density threshold (of the order of $10^{11}$ g/cm$^3$)  and that are expected 
to behave as a superfluid except at very high temperature, and on the other 
hand the protons, which (together with the ions to which they are confined) 
will behave as an effectively ``normal'' fluid in the crust, though they may 
behave as another superfluid in the deeper layers at densities above the 
nuclear threshold of the order of $10^{14}$ g/cm$^3$ at which the ions merge.

One of the main goals of this article is to clarify the relationships
between different kinds \cite{Mendell91,Prix01} of definition of 
``effective mass'' that have been introduced by various authors as a 
quantitative measure of the  ``entrainment'' effect, and particularly to 
provide a discussion of the way such masses depend on the gauge freedom 
involved in the choice of a chemical basis for the purpose of specifying 
just which neutrons are considered to be free. In the homogeneous 
(unclustered) phase above the nuclear density threshold prevailing in the 
liquid core the neutrons will all be effectively free, while in the outer 
crust below the ``drip'' threshold it is clear that none of them will be 
free, in the ``operational'' sense of being able to penetrate the potential 
barriers between nuclei in an astrophysical timescale shorter than the age 
of the universe. In the inner crust at densities just above this threshold, 
neutrons in low and intermediate energy states will still be effectively 
confined, but there will be a clear cut critical energy above which they 
will be effectively free to travel over ionic separation distances on a 
microscopically short timescale.

The problem is that, at deeper levels in the crust, such an ``operational''
discrimination between ``confined'' and ``free'' neutron states  will no 
longer be so clear cut, because the ionic wells will get too close for the 
exponential suppression outside to be fully effective, so that there will 
be marginally bound states with intermediate penetration timescales that are
macroscopically long but cosmologicaly short. This means that there will be
some degree of ambiguity that needs to be resolved by a more or less
arbitrary choice of a chemical gauge whereby the baryon number 
{\bb \AA\ \fb} of the ionic clusters is prescribed as a function of their
(depth dependent) charge number {\bb\ZZ\fb}. Fortunately, as will be shown 
below, the most important quantities in the two fluid formalism at a local 
(intervortex) level notably the action density {\bb \Lambda \fb} with its
concomitant energy function, and in particular the (irrotational)
superfluid neutron momentum, all turn out to be robustly independent of 
which particular (``operational'' or other) gauge criterion is used.

Following this discussion of the different ways of formulating the generic
treatment of entrainment in the two fluid formalism, a more particular
purpose of this work is to show how quantitative values of the relevant 
``effective mass'' are provided by the corresponding (gauge independent) 
``mobility coefficient'' {\bb\calK\fb} for which quantitative estimates are
obtainable from the kind of microscopic analysis described in our 
immediately preceding work~\cite{CCHI}. This analysis is based on a 
non-relativistic quantum mechanical formalism of the type introduced by 
Oyamatsu and Yamada~\cite{Oyamatsu94} in a local frame with respect to which 
the ionic crust lattice is at rest, so that it gives rise to a static 
effective potential having a periodic dependence on the cartesian space 
coordinates {\bb x^{ i}\fb}. The analysis is carried out in terms of single 
particle wave functions subject to Bloch type boundary conditions (meaning 
that they deviate from ordinary periodicity by a phase factor of the form
{\bb \exp\{{\rm i} \kk_i x^{ i}\}\fb}) of the kind that is commonly used
for electrons in ordinary solid state physics, but that has not yet been 
systematically employed for analogous problems involving neutrons. In a 
follow up article~\cite{CCHIII} we have shown how the method can be 
extended to include the moderate adjustments arising from the effect of BCS 
type superfluid pairing.

The mobility coefficient  {\bb{\calK}\fb}, that is provided by this 
analysis, has a value which is interpretable as the ratio of the density 
{\bb \nn\fb} of ``free'' (``conduction'') neutrons to their effective mass 
{\bb \mm_\star\fb}. It was suggested by our preliminary results~\cite{CCHI}  
and has been confirmed by more recent work ~\cite{Chamel04} that this value 
is likely to be much larger than had been expected. Even when measured with 
respect to an ``operational'' gauge as shown in  Figure \ref{effmass_fc}
in which the number density {\bb \nn\fb} of neutrons that are counted as 
``free'' allows only for those that are not bound inside ionic nuclei, the 
quantity {\bb \mm_\star\fb} is likely to become extremely large compared
with the ordinary neutron mass {\bb\mm\fb} in the middle layers of the
inner crust. It would of course be even larger with respect to the very
simple ``comprehensive'' gauge in which all the neutrons are counted as
``free'' so that, as shown in Figure \ref{effmass_np}, the corresponding
``effective mass'' would in any case diverge as the density decreases to
the ``neutron drip'' value.

The purely Newtonian (i.e. non relativistic) framework within which the
work of the present article is carried out should be adequate, on a local
scale, as a fairly good approximation in the density regime under
consideration. This lower crust regime extends from the ``neutron drip''
threshold to the transition to a purely fluid (at low temperatures actually
superfluid) mixture of neutrons and protons  at the base of the crust where
the mean density approaches that of ordinary nuclear matter (at a value
of the order of about $10^{14}\, {\rm g.cm}^{-3}.$) In a global
representation of the neutron star (on scales of the order of a kilometer
or more) there will be significant deviations from flat geometry whose
quantitative treatment would require the use of General Relativity, but
the discussion here will be limited to a local neighbourhood that is
sufficiently small (on scales of the order of a centimeter or less) to be
treated as approximately homogeneous and geometrically flat. Even on such
a purely local scale it would be necessary to use a special relativistic
(Minkowski space) description if we were concerned with effects involving
the electrons, but the rest mass of the neutrons, to which the present
analysis is restricted, is so much higher that a purely Newtonian analysis
should be quite adequate in the moderate energy range (up to a few tens of
MeV) that is involved. This contrasts with the state of affairs in the fluid 
core of the neutron star, where estimation of the relevant ``entrainment'' 
between (superfluid) neutrons and protons requires a different kind of 
treatment~\cite{ComerJoynt03} in which allowance for relativity is 
important, but where the effect of the entrainment is much more moderate. 
An implication is that to improve on previous work~\cite{CarSedLan} that 
ignored the effects both of relativity and of entrainment in the analysis
of angular momentum transfer in the crust, the effect of relativity is of
secondary importance, and that the first priority  is
to include the relevant allowance for the effective mass enhancement due 
to entrainment~\cite{CarCham05}.

\section{Generic two-fluid models with stratification}
\subsection{The canonical master function}

In astrophysical contexts for which a non-relativisitic Newtonian 
mechanical description is adequate it will usually be sufficiently
accurate to represent the relevant distribution of mass just in terms 
of a conserved current of baryons characterised by a single constant mass 
parameter,  $\mm$ say, that can be taken to be the standard atomic mass
unit or simply the proton mass which is different only by a fraction of 
a per cent and so is near enough for practical purpose. It is 
convenient for many purposes to use a 4-dimensional notation scheme
\cite{CC03,CC04,CC05} in which, for example, the total baryon number
density {\bb \nn_{\rm b}\fb} say, and the corresponding baryon current 
density components {\bb\nn_{\rmb}^{\,i}\fb} with respect to space
coordinates {\bb x^i\fb} {\bb (i=1,2,3)\fb} are combined to form a 4-current
{\bb \nn_{\rmb}^{\,\nu}\fb} {\bb(\nu=0,1,2,3)\fb} whose time component 
is given by {\bb \nn_{\rmb}^{^{_0}}=\nn_{\rmb}\fb}. This means that the
total baryon conservation law,
{\be {\partial \nn_{\rmb}/\partial t}+\nabla_{\!i\,}
\nn_{\rmb}^{\,i}=0\, ,\label{40}\fe}
will be expressible more concisely  as
{\be \nabla_{\!\nu}\nn_{\rmb}^{\,\nu}=0\, .\label{41}\fe} 
(Throughout this paper, summation over repeated indices is assumed).

Our purpose here is to consider situations involving relative
motion between an uncharged contribution {\bb\nn_{\rmn}^{\,\nu}\fb}
that will be attributable mainly to neutrons (but perhaps including 
other neutral hyperons such as the {\bb\Lambda\fb} in the inner core 
regions) and an electrically charged contribution 
{\bb\nn_{\rmp}^{\,\nu}\fb} that will similarly be attributable
mainly to protons. (It is to be understood that overall electrical
charge balance is ensured by a non baryonic current attributable
mainly to electrons, whose mass contribution is so small that it
can be neglected.) The two independent baryonic contributions 
combine to give the conserved total baryon current
{\be \nn_{\rmb}^{\,\nu}=\nn_{\rmn}^{\,\nu}+\nn_{\rmp}^{\,\nu}
\, .\label{44}\fe}
In dynamic processes on short timescales the neutron and proton
currents may each be considered to be separately conserved, meaning 
that the 4-divergences {\bb \nabla_{\!\nu}\nn_{\rmn}^{\,\nu}\fb} and
{\bb \nabla_{\!\nu}\nn_{\rmp}^{\,\nu}\fb} will both vanish individually, 
but in long term  ``secular'' evolution processes it will be
necessary to allow for the possibility of converting neutrons to
protons or vice versa by weak interaction processes.

The present discussion will be concerned with cases in which the
evolution of the separate contributions {\bb\nn_{\rmn}^{\,\nu}\fb} and 
{\bb\nn_{\rmp}^{\,\nu}\fb} can be described  by a multiconstituent fluid 
model of the kind governed by a master function {\bb\Lamb_{\rm mat}\fb}, 
that acts as the material (meaning non gravitational) part of a 
Lagrangian in effectively conservative applications \cite{CC03,CC04}, 
but that is also compatible with allowance for potentially relevant 
dissipative effects~\cite{CC05}. Such a multifluid description does 
however entail neglect of the extra energy contributions due to possible 
elastic solid deformations \cite{CC06}, and to frozen in magnetic 
fields~\cite{CCC05}, whose inclusion would require a more elaborate 
treatment. As well as depending on the separate currents 
{\bb\nn_{\rmn}^{\,\nu}\fb} and {\bb\nn_{\rmp}^{\,\nu}\fb} it is useful 
to allow for the potentially important effect of stratification due 
to baryonic clustering forming ionic nuclei characterised by a charge 
number {\bb \ZZ\fb} and hence by a number density that is expressible as
{\be \nn_{\rmI}=\nn_{\rmp}/\ZZ\, ,\label{Ab}\fe}
wherever the density is below the saturation density of the
order of $10^{14}$ g.cm$^3$. This means that the generic variation of the 
master function will be given in terms of corresponding partial 
derivative coefficients by an expression of the form 
{\be \delta \Lamb_{\rm mat}= -\mmu^{\rmI}\,\delta\nn_{\rmI}
+\mmu^{\rmn}_{\,\nu}\,\delta 
\nn_{\rmn}^{\,\nu}+\mmu^{\rmp}_{\,\nu}\,\delta \nn_{\rmp}^{\,\nu}
\, ,\label{51}\fe}
or equivalently in the less compact form of a traditional 3+1 space
time decomposition by an expression of the form
{\be \delta \Lamb_{\rm mat}=-\mmu^{\rmI}\,\delta\nn_{\rmI}
- \mmu^{\rmn}\,\delta \nn_{\rmn}-
\mmu^{\rmp}\,\delta \nn_{\rmp}+\mmu^{\rmn}_{\,i}\,\delta \nn_{\rmn}^{\,i}+
\mmu^{\rmp}_{\,i}\,\delta \nn_{\rmp}^{\,i}\, ,\label{52}\fe}
in which {\bb \mmu^{\rmn}_i\fb} and {\bb\mmu^{\rmp}_i\fb} will be
respectively interpretable as the mean momentum per particle of the
neutrons and of the protons, while the quantities
{\be \mmu^{\rmn}=-\mmu^{\rmn}_{\,_0}\, ,\hskip 1 cm
 \mmu^{\rmn}=-\mmu^{\rmn}_{\,_0}\, ,\fe}
will be interpretable as the corresponding neutronic and protonic chemical 
potentials, while the coefficient {\bb\mmu^{\rmI}\fb} is an ionic cluster
potential, whose gradient, if any, represents the effect of stratification. 
This potential can be used to construct a corresponding 4-momentum
covector that is given in terms of the gradient of the 
Newtonian time coordinate {\bb x^{_0}=t \fb} by the formula
{\be \mmu^{\rmI}_{\,\nu}=-\mmu^{\rmI}\, t_\nu\, ,\hskip 1 cm t_\nu=
\nabla_{\!\nu}\, t=\delta^{_0}_{\,\nu} \, ,\fe}
which means that it  has vanishing space components, {\bb \mmu^{\rmI}_{\,i}
=0 ,\fb} as an expression of the fact that (in the Newtonian limit)
the clustering contributes only to energy but not to mass.

The models we are considering are intended for application to neutron star 
matter after it has cooled down sufficiently for its dynamics to be hardly 
affected by thermal effects, which are not included. These models are to 
be interpreted as being governed by dynamical equations that are formally 
identical to those of the heat convecting thermal model presented in the 
appendix of the work~\cite{CC05} referred to above,  with the ionic
cluster potential {\bb\mmu^{\rmI}\fb} taking the place of the temperature 
{\bb\Thet\ ,\fb}  and with the ionic number density {\bb\nn_{\rmI}\fb} 
taking the place of the entropy density {\bb \ss\fb}, so that the entropy
current 4-vector {\bb\ss^\mu\fb} is to be replaced here by the ionic number 
current 4-vector {\bb\nn_{\rmI}^{\,\nu}\fb} as given 
in terms of the flow 4-vector {\bb \uu_\rmc^{\, \nu}\fb} of the crust by
{\be \nn_{\rmI}^{\,\nu}=\nn_{\rmI}\, \uu_\rmc^{\,\nu}\, ,\hskip 1 cm
\uu_\rmc^{\,\nu}=\nn_{\rmp}^{\,\nu}/ \nn_{\rmp}\, \, .\fe}

In terms of the generalised pressure function
{\be \Pssi=\Lamb_{\rm mat}+\nn_{\rmI}\,\mmu^{\rmI}-
\nn_{\rmp}^{\, \nu}\mmu^{\rmp}_{\,\nu}-
\nn_{\rmn}^{\, \nu}\mmu^{\rmn}_{\,\nu}\, ,\fe}
and the Newtonian gravitational potential {\bb \phi\fb} the
corresponding material energy momentum 4-tensor~\cite{CC04} will
therefore be given by
{\be \TT^\mu_{\ \, \nu}=\nn_{\rmI}^{\, \mu}\mmu^{\rmI}_{\,\nu}
+\nn_{\rmp}^{\, \mu}\mmu^{\rmp}_{\,\nu}+
\nn_{\rmn}^{\, \mu}\mmu^{\rmn}_{\,\nu}+\Pssi\delta^\mu_{\,\nu}
-\phi\,\mm\nn_{\rmb}^{\,\mu} t_\nu\, .\label{enmom}\fe}
This means that, if the only external force is gravitational, the total
force balance equation will then be expressible as
{\be \nabla_{\!\mu}\TT^\mu_{\ \, \nu}=-\mm\nn_\rmb
\nabla_{\!\nu}\phi\, ,\label{forcebal}\fe}
while the condition of conservation of entropy in the conservative 
case is to be replaced here by the ionic cluster conservation condition
{\be \nabla_{\!\nu}\nn_{\rmI}^{\,\nu}=0\, ,\label{ioncon}\fe}
which can be expected to be valid, not just in short timescale dynamic 
processes in which there would be separate conservation of the
neutrons
{\be \nabla_{\!\nu}\nn_{\rmn}^{\, \nu}=0\label{neucon}\fe}
(and hence also, according to (\ref{41}),  of the protons) but even in 
processes having the much longer timescales needed for weak interactions
to adjust the relative numbers of protons and neutrons so as to achieve 
the condition of ``beta  equilibrium'', namely vanishing of the affinity
{\be  \calA=(\mmu^{\rmn}_\nu-\mmu^{\rmp}_\nu) 
\uu_\rmc^{\,\nu}\, ,\label{affin}\fe}
as specified with respect to the ``normal'' rest frame
specified by {\bb \uu_\rmc^{\,\nu}\fb}, which is that of the 
protons (and therefore also of the ions) as opposed to that of the neutrons,
whose superfluidity at low temperatures allows them to retain a relative 
motion (at densities above the neutron drip threshold) even in a state of
exact non dissipative equilibrium.  On timescales even longer than
needed for the beta equilibrium condition,
{\be \calA=0\, ,\label{beteq}\fe}
it is ultimately to be expected that barrier tunnelling processes involving 
nuclear fission and recombination would lead to violation of (\ref{ioncon}) 
in such a way as to efface the stratification by allowing the ionic cluster 
potential {\bb\mmu^{\rmI}\fb} to tend towards zero, but in practice the
stratification will commonly survive long enough to have important 
stabilising consequences, particularly in the outer layers of the neutron 
star.

To provide a complete system of evolution equations for the dynamic
variables {\bb \nn_\rmI,\fb} {\bb \nn_\rmn^{\,\nu},\fb} and
 {\bb \nn_\rmp^{\,\nu},\fb} the system consisting of (\ref{forcebal})
and (\ref{ioncon}), together with one of the alternative
possibilities (\ref{neucon}) or (\ref{beteq}), does not suffice, but
must be supplemented by a further condition which can simply be taken to be
the mesoscopic neutron superfluidity condition to the effect that the total 
neutron momentum covector,
{\be \ppi^{\rmn}_{\,\nu}=\mmu^\rmn_{\,\nu}-\phi\,\mm t_\nu
\, ,\label{pif}\fe}
should be proportional to a phase gradient, taking the form
{\be \ppi^{\rmn}_{\,\nu}=\hba\nabla_{\!\nu\,} 
\varphi^\rmn\, ,\label{superf}\fe}
in which the phase variable {\bb \varphi^\rmn\fb}
has period  {\bb\pi\fb} (not {\bb2\pi\fb}) due to the fermionic half integer
spin of the neutrons \cite{CCHIII}. (This corresponds to a period having the usual 
value {\bb2\pi\fb} for the corresponding phase {\bb\varphi=2\varphi^\rmn\fb}
for Cooper pairs with momentum covector {\bb 2\ppi^{\rmn}_{\,\nu} \fb}.)

\subsection{Chemical invariance of superfluid momentum}
\label{chemin}

In order to obtain a formulation that is consistent with traditional 
usage in the low density regime with density less than the order of
$10^{11}$ g.cm$^3$ for which the neutrons will also be effectively 
confined within the nuclei, it is useful to introduce a dependent variable, 
{\bb\AA_\rmc\fb}, that is to be interpreted as  the total number of 
effectively ``confined'' baryons per nucleus, and that specifies a 
corresponding ``confined'' baryon number current 
{\bb \nn_{\rmc}^{\,\nu} \fb}, which will be given by
{\be \nn_{\rmc}^{\,\nu}=\AA_\rmc \nn_{\rmI}^{\,\nu}\, .\label{confdef}\fe}
The``confined'' baryon number {\bb \AA_\rmc=\nn_\rmc/\nn_\rmI\fb} will
evidently lie somewhere in the range   {\bb \ZZ\leq\AA_\rmc\leq\AA_\rmb\fb}
where {\bb \AA_\rmb=\nn_\rmb/\nn_\rmI\fb} is the total baryon number 
per nucleus, which will of course be the same as {\bb\AA_\rmc\fb}
at densities below the ``neutron drip'' threshold. As a corollary there
will also be a corresponding ``free'' neutron current given by
{\be \nn_{\rmf}^{\,\nu}=
 \nn_{\rmb}^{\,\nu}- \nn_{\rmc}^{\,\nu}\, ,\label{freedef}\fe}
which will be non vanishing at densities above the ``neutron drip'' 
threshold. In the deeper layers of the ionic ``crust''  region where the 
clusters are near to stage of merging they may become highly deformed to 
``spaghetti'' or even ``lasagna'' shaped configurations \cite{PR95, Hae01} 
for which (as remarked above in the introduction) the meaning of the 
absolute numbers {\bb \ZZ\fb} and {\bb\AA\fb} may become rather hazy, but 
this need not prevent the ratio
{\be \Alf_\rmc=\AA_\rmc/\ZZ=\nn_{\rmc}/\nn_{\rmp}\fe}
of the ``confined'' baryon number density to the charged baryon number
density from remaining definable according to some ansatz that might 
reasonably be required to specify {\bb \Alf_\rmc\fb} as a weakly dependent
function of {\bb \ZZ\fb} in such a way that beyond the transition to an
unclustered homogeneous phase above ordinary nuclear matter density, the 
corresponding ``free'' neutron current {\bb \nn_{\rmf}^{\,\nu}\fb} will 
simply become the same as the entire neutron current  
{\bb \nn_{\rmn}^{\,\nu}\fb}, while at lower densities it would be 
determined by the equation of state for ``cold catalysed'' matter whereby
all relevant scalars including {\bb\ZZ\fb} and {\bb\AA_\rmc\fb} are given 
(in such a way as to minimise the energy density) as functions of {
\bb\nn_\rmb\fb}.

However that may be, subject to the specification of any reasonable 
(realistic or idealised) equation of state for  {\bb \AA_\rmc\fb} as a 
function just of {\bb \ZZ\fb} so that the variation of the ratio 
{\bb \Alf_\rmc\{\ZZ\}\fb} will be given in terms of its derivative
 {\bb\Alf_\rmc^\prime={\rm d}\Alf_\rmc/{\rm d}\ZZ\fb} by an expression of 
the form
{\be \delta\Alf_\rmc=\ZZ\Alf_\rmc^\prime\left(\frac{\delta \nn_\rmp}
{\nn_\rmp}-\frac{\delta \nn_\rmI}{\nn_\rmI}\right)\, ,\fe}
it will be possible, and for some purposes convenient, to make a 
corresponding chemical basis transformation whereby the new (empirically 
defined) current variables
 {\bb\nn_{\rmf}^{\,\nu}\fb} and {\bb\nn_{\rmc}^{\,\nu}\fb} are used in
place of the original (more physically fundamental) current variables
{\bb\nn_{\rmn}^{\,\nu}\fb} and {\bb\nn_{\rmp}^{\,\nu}\fb} as the
independent variables of the system. In terms of the new chemical basis, 
as specified by the homogeneous linear transformation
{\be \nn_{\rmf}^{\,\nu}=\nn_{\rmn}^{\,\nu}+(1-\Alf_\rmc)\nn_{\rmp}^{\,\nu}
\, ,\hskip 1 cm  \nn_{\rmc}^{\,\nu}=\Alf_\rmc\nn_{\rmp}^{\,\nu}
\label{chemtrans}\, ,\fe}
the generic variation (\ref{51}) of the master function and the 
corresponding energy momentum tensor (\ref{enmom}) will be expressible in 
the exactly analogous forms
{\be \delta \Lamb_{\rm mat}= -\tilde\mmu^{\rmI}\,\delta\nn_{\rmI}
+\mmu^{\rmf}_\nu\,\delta 
\nn_{\rmf}^{\,\nu}+\mmu^{\rmc}_\nu\,\delta \nn_{\rmc}^{\,\nu}
\, ,\label{deLamb}\fe}
and
{\be \TT^\mu_{\ \, \nu}=\nn_{\rmI}^{\, \mu}\tilde\mmu^{\rmI}_{\,\nu}
+\nn_{\rmc}^{\, \mu}\mmu^{\rmc}_{\,\nu}+
\nn_{\rmf}^{\, \mu}\mmu^{\rmf}_{\,\nu}+\Pssi\delta^\mu_{\,\nu}
-\phi\,\mm\nn_{\rmb}^{\,\mu} t_\nu\, ,\label{enmomcom}\fe}
with
{\be \Pssi=\Lamb_{\rm mat}+\nn_{\rmI}\,\tilde\mmu^{\rmI}-
\nn_{\rmc}^{\, \nu}\mmu^{\rmc}_{\,\nu}-
\nn_{\rmf}^{\, \nu}\mmu^{\rmf}_{\,\nu}\, ,\label{presf}\fe}
in which the adjusted ionic cluster potential {\bb\tilde\mmu^{\rmI}\fb}
and the correspondingly adjusted analogue
{\be\tilde\calA=(\mmu^{\rmf}_\nu-\mmu^{\rmc}_\nu) \uu_\rmc^{\,\nu}
\,  ,\fe}
of the affinity (\ref{affin}) will be related to their untransformed
analogues  by 
{\be \calA =\tilde\calA(\Alf_\rmc+\ZZ\Alf_\rmc^\prime)\label{adjaf}\fe} 
and
{\be \mmu^{\rmI}=\tilde\mmu^{\rmI}-\ZZ^2\tilde\calA\,\Alf_\rmc^\prime
\, ,\fe}
while the new ``free'' and ``confined'' baryon momentum 4 covectors
 {\bb\mmu^{\rmf}_\nu \fb} and  {\bb\mmu^{\rmc}_\nu \fb} will be given in
terms of their untransformed analogues  {\bb\mmu^{\rmI} ,\fb} 
{\bb\mmu^{\rmf}_\nu , \fb} {\bb\mmu^{\rmc}_\nu , \fb} by the 
relations
{\be \mmu^{\rmp}_{\,\nu}= \mmu^{\rmc}_{\, \nu}+(1-\Alf_\rmc)
( \mmu^{\rmf}_{\,\nu}- \mmu^{\rmc}_{\,\nu})-\ZZ\tilde\calA\,\Alf_\rmc^\prime
t_\nu\,\fe}
and
{\be \mmu^{\rmn}_{\,\nu}= \mmu^{\rmf}_{\, \nu}\, .\fe}
This last result is important: it tells us that whereas the momentum
covector {\bb \mmu^{\rmc}_{\, \nu}\fb} associated with the ``confined'' 
part has a rather complicated dependence on the chemical gauge 
specified by the choice of the functional dependence of 
{\bb\Alf_\rmc\fb} on {\bb\ZZ ,\fb} on the other hand the momentum 
covector {\bb\mmu^{\rmf}_{\, \nu}\fb} associated with the ``free'' 
neutron current is invariant with respect to the change of gauge. The 
correponding total
{\be \ppi^{\rmf}_{\,\nu}=\mmu^\rmf_{\,\nu}-\phi\,\mm t_\mu
\, ,\fe}
is just the same as its analogue (\ref{pif}) in the original
formulation, so that the superfluidity condition (\ref{superf})
will be expressible in terms of the
the same phase variable {\bb\varphi\fb} in exactly the same
way as before, meaning that it will be given simply by
{\be \ppi^{\rmf}_{\,\nu}=\hba\nabla_{\!\nu\,}
\varphi\, .\fe}

The upshot of this is that if the set of dynamical equations is
completed by the beta equilibrium condition (\ref{beteq}), which
according to (\ref{adjaf}) will be expressible in the language of the 
new formulation simply by
{\be\tilde\calA=0\, ,\fe}
then the entire system will be chemically covariant in the sense
of having the same form regardless of the choice of the gauge function
{\bb \Alf_\rmc\{\ZZ\}\fb} in the chemical basis transformation
(\ref{chemtrans}).

On the other hand if (as will be appropriate for cases involving
relatively short dynamical timescales) the beta equilibrium condition 
(\ref{beteq}) is replaced by the separate neutron current conservation 
condition (\ref{neucon}) then the complete system will be chemically 
covariant only with respect to transformations in which 
{\bb\Alf_\rmc\fb} is simply taken to be constant, as for example in what 
we shall refer to as the ``comprehensive gauge'', which is given simply
by {\bb\Alf_\rmc=1\fb} so that absolutely all of the neutrons are 
classified as ``free'', while a less trivial example is that of what may 
be described as the ``paired gauge'', which is given by  
{\bb\Alf_\rmc=2\fb}, and which is interpretable as meaning that the only
 neutrons classified as ``confined''  are those in the tightly bound
states that are directly paired with  corresponding proton states in the
nuclei. However,  if we wish to use {\bb\Alf_\rmc\fb} to provide a more 
realisticlly ``operational'' estimate of the total baryon to proton number
ratio in the nuclei, it will need to be given a non vanishing derivative 
{\bb\Alf_\rmc^\prime ,\fb} and in that case the corresponding formulation 
of the separate conservation condition \rf{neucon}\fr will have the 
patently gauge dependent form
{\be \nabla_{\!\nu\,}\nn_\rmf^{\,\nu}=-\Alf_\rmc^\prime\nn_\rmc^{\,\nu}
\nabla_{\!\nu\,}\ZZ\, .\fe}

\subsection{Local energy function}

In order to obtain information about the equation of state giving the 
functional dependence of the material action density {\bb\Lamb_{\rm mat}\fb} 
on the relevant independent variables by comparison with the results of a 
microphysical analysis, it is convenient to work with the corresponding 
energy function. The complete energy momentum tensor \rf{enmomcom}\fr 
comes from a complete action density {\bb\Lamb=\Lamb_{\rm mat}+
\Lamb_{\rm pot}\fb} in which, as discussed in the previously cited work 
\cite{CC04}, the gravitational potential energy contribution is given simply 
by {\bb\Lamb_{\rm pot}=-\rrho\phi ,\fb} with {\bb\rrho=\mm\nn_\rmb\fb}
so the corresponding energy {\bb\UU=-\TT^{_0}_{\ _0}\fb} will have the form
of a sum {\bb\UU=\UU_{\rm mat}+\UU_{\rm pot}\fb} involving a 
gravitational contribution {\bb\UU_{\rm pot}=-\Lamb_{\rm pot}\fb} that 
must be subtracted off to leave the purely material part 
{\bb\UU_{\rm mat}\fb}  corresponding just to  {\bb \Lamb_{\rm mat}\fb}. It 
can thus be seen that this material energy density will be given (for an 
arbitrary choice of the chemical gauge function {\bb\Alf_\rmc\{\ZZ\}\fb}) 
by
{\be \UU_{\rm mat}=\tilde\mmu^{\rmI}\,\nn_\rmI -
\mmu^{\rmf}_{_0}\, \nn_{\rmf}-\mmu^{\rmc}_{_0}\, \nn_{\rmc}
-\Pssi  \, ,\label{54}\fe}
which, by \rf{presf}\fr, is evidently equivalent to taking
{\be \UU_{\rm mat} =\mmu^{\rmf}_i\, \nn_{\rmf}^{\,i}+
\mmu^{\rmc}_i\, \nn_{\rmc}^{\,i} -\Lamb_{\rm mat}\, .\label{55}\fe}

So long as the relevant velocities are sufficiently small,  it will be
possible to decompose the material Lagrangian and energy density in the 
form
{\be \Lamb_{\rm mat}=\Lamb_{\rm ins}+\Lamb_{\rm dyn} \, ,\label{56}\fe}
and
{\be \UU_{\rm mat}=\UU_{\rm ins}+\UU_{\rm dyn} \, ,\label{57}\fe}
in terms of a static internal energy contribution
{\be \UU_{\rm ins}=-\Lamb_{\rm ins}\, ,\label{58}\fe}
that will depend only on the relevant ionic and (free and confined) 
baryon number densities namely {\bb \nn_{\rmI} ,\fb} 
{\bb \nn_{\rmf} ,\fb} {\bb \nn_{\rmc}\fb}, together with a dynamic 
contribution for which the velocity dependence is homogeneously 
quadratic so that as for the purely kinetic part \cite{CC04} the 
corresponding energy contribution will be the same as the Lagrangian 
contribution from which it is derived, having the form
{\be \Lamb_{\rm dyn} =\UU_{\rm dyn}
=\frac{_1}{^2}( \mmu^{\rmf}_i\, \nn_{\rmf}^{\,i}+
\mmu^{\rmc}_i\, \nn_{\rmc}^{\,i})\, .\label{dynamic}\fe} 
This means that it will be given as a function of the current components
{\bb \nn_{\rmf}^{\,i}\fb} and {\bb \nn_{\rmc}^{\,i}\, \fb} by an 
expression of the form
{\be \Lamb_{\rm dyn}=\frac{_1}{^2}\gamma_{ij}
\big({\calK}^{\rm ff}\,
\nn_{\rmf}^{\,i}\,\nn_{\rmf}^{\,j}+2{\calK}^{\rmfc}\,\nn_{\rmf}^{\,i}\,
\nn_{\rmc}^{\,j}+{\calK}^{\rm cc}\,\nn_{\rm c}^{\,i}
\,\nn_{\rmc}^{\,j}\big)\, ,\label{59}\fe}
in which the coefficients {\bb \calK^{\rm ff}\fb},
{\bb \calK^{\rm cc}\fb}, and {\bb \calK^{\rmfc}\fb} are functions only
of {\bb \nn_{\rmI} ,\fb} {\bb \nn_{\rmf} ,\fb} and {\bb \nn_{\rmc}\,\fb}.
It follows from the Galilean covariance requirements \cite{CC03} in the
the decomposition \rf{67}\fr to be discussed in the
next section that these coefficients must satisfy the conditions
{\be \calK^{\rm ff}\nn_{\rmf} +  \calK^{\rm fc} \nn_{\rmc} = \mm 
\, ,\hskip 1 cm
\calK^{\rm cc}\nn_{\rmc} +  \calK^{\rm fc} \nn_{\rmf} = \mm
\, ,\label{gal1}\fe}
which means that only one of these coefficients needs to be specified.
It follows from \rf{59}\fr that the momenta defined by \rf{52}\fr will be 
given as homogeneous linear functions of the currents by
{\be \mmu^{\rmf}_{\,i}=\gamma_{ij}\big({\calK}^{\rm ff}\,\nn_{\rm f}^{\,j}
+{\calK}^{\rmfc}\, \nn_{\rmc}^{\,j}\big)\, ,\hskip 1 cm
\mmu^{\rmc}_{\,i}=\gamma_{ij}\big({\calK}^{\rmfc}\,\nn_{\rmf}^{\,j}
+{\calK}^{\rm cc}\,n_{\rmc}^{\,j}\big)\, .\label{61}\fe}
These relations will be invertible to give the expressions
{\be \nn_{\rmf}^{\,i}=\gamma^{ij}\big({\calK}_{\rm ff}\,\mmu^{\rmf}_{\,j}
+{\calK}_{\rmfc}\,\mmu^{\rmc}_{\,j}\big)\, ,\hskip 1 cm
\nn_{\rmc}^{\,i}=\gamma^{ij}\big({\calK}_{\rmfc}\,\mmu^{\rmf}_{\,j}
+{\calK}_{\rm cc}\,\mmu^{\rmc}_{\,j}\big)\, ,\label{60}\fe}
in which the original coefficients will be related to the new ones by
{\be {\calK}_{\rm ff}\!=\! \frac{ {\calK}^{\rm cc}}{
{\calK}^{\rm ff}{\calK}^{\rm cc}\!-\!{\calK}^{\rmfc\, 2}}\, ,\hskip
0.3 cm {\calK}_{\rmfc}\!=\! \frac{ -{\calK}^{\rmfc}}{{\calK}^{\rm ff}
{\calK}^{\rm cc}\!-\!{\calK}^{\rmfc\,2}}\, ,\hskip 0.3 cm
{\calK}_{\rm cc}\!=\! \frac{ {\calK}^{\rm ff}}{ {\calK}^{\rm ff}
{\calK}^{\rm cc}\!-\!{\calK}^{\rmfc\, 2} }\, .\label{62}\fe}
In terms of these new coefficients, the covariance conditions \rf{gal2}\fr
are expressible as
{\be \calK_{\rm ff} +  \calK_{\rm fc}= \frac{\nn_{\rmf}}{\mm} 
\, ,\hskip 1 cm
\calK_{\rm cc}+  \calK_{\rm fc} = \frac{\nn_{\rmc}}{\mm}
\, .\label{gal2}\fe}
The expression \rf{59}\fr is convertible to the dual form
{\be \UU_{\rm dyn}=\frac{_1}{^2}\gamma^{ij}\big({\calK}_{\rm ff}\,
\mmu^{\rmf}_{\,i}\,\mmu^{\rmf}_{\,j}+2{\calK}_{\rmfc}\,
\mmu^{\rmf}_{\,i}\,\mmu^{\rmc}_{\,j}+{\calK}_{\rm cc}\,
\mmu^{\rmc}_{\,i}\,\mmu^{\rmc}_{\,j}\big)\, .\label{59bis}\fe}

The requirement that (in order to ensure that the minimum energy
configuration is the one in which the currents vanish) the dynamical energy
density {\bb \UU_{\rm dyn} \fb} in expressions \rf{59}\fr or \rf{59bis}\fr
should be positive definite , implies that the on-diagonal coefficients
should all be positive, as well as the restrictions
{\be {\calK}_{\rm ff}\,{\calK}_{\rm cc} > ({\calK}_{\rmfc})^2, \hskip 1 cm
{\calK}^{\rm ff}\,{\calK}^{\rm cc} > ({\calK}^{\rmfc})^2 \label{64bis}
\, ,\fe}
which can be seen from \rf{62}\fr to entail that the off-diagonal
coefficients {\bb {\calK}^{\rmfc} \fb} and {\bb {\calK}_{\rmfc} \fb} 
should  have opposite signs.

\subsection{Specification of the entrainment coefficient}

In order to relate the canonical approach described above to a description 
of the traditional kind~\cite{Mendell91, Andreev76} in terms of 3-velocities 
and densities, the dynamical energy density has to be rewritten in terms of 
the relevant current 3-velocity vectors {\bb v_\rmc^{\,i}\fb} and 
{\bb v_\rmf^{\,i}\fb} as specifed with respect to chosen Galilean rest 
frame by setting
{\be \nn_\rmc^{\,i}=\nn_\rmc\, v_\rmc^{\,i}\, ,\hskip 1 cm
 \nn_\rmf^{\,i}=\nn_\rmf\, v_\rmf^{\,i}\ ,\label{46}\fe}
which means that {\bb v_\rmc^{\,i}\fb} will be physically well defined as
the  ``normal'' velocity of the ions and the rigorously confined proton 
current, for which we shall have 
{\be \nn_{_{\rmI}}^{\,i}= \nn_{_{\rmI}}\, v_{\rmc}^{\,i}\, ,\hskip 1 cm
\nn_{\rmp}^{\,i}= \nn_{\rmp}\, v_{\rmc}^{\,i}\, .\label{49}\fe}
whereas {\bb v_\rmf^{\,i}\fb} depends on the chemical gauge function 
{\bb\Alf_\rmc\{\ZZ\}\fb} that is chosen to determine which of the neutrons 
are classified as ``free''. The chemical gauge dependent but  Galilean frame 
independent relative flow velocity vector
{\be \bar v_{\rmfc}^{\,i}=v_{\rmf}^{\,i}-v_{\rmc}^{\,i}\, ,\label{47}\fe}
can then be used for construction of the relative current 3-vector
{\be \nn^i=\nn_\rmf\,  \bar v_{\rmfc}^{\,i}\, ,\label{48}\fe}
which as well as being evidently frame independent, also has the
noteworthy property of being physically well defined in the sense that
(like the neutron momentum covector {\bb \mmu^\rmf_{\,i}\fb})  it is
unaffected by a change of the gauge parameter {\bb\Alf_\rmc\fb} in
\rf{chemtrans}\fr. This can be seen from the possibility of rewriting it 
directly in terms of the unambiguously well defined total baryonic number 
density {\bb \nn_{\rmb}=\nn_{\rmf}+\nn_{\rmc}\fb}, and the 
unambigously defined ``normal'' 3-velocity vector {\bb v_{\rmc}^{\,i}\fb}, 
in the form
{\be \nn^i=\nn_{\rm b}^{\,i}-\nn_{\rm b} v_{\rmc}^{\,i}\, ,\label{50}\fe}
which is manifestly independent of any prescription for specifying which
 neutrons are deemed to be ``confined'' and which are deemed to be ``free''. 
(Before continuing it is to be remarked that subject to substitution of 
total lepton number current, {\bb \nn_{\rm e}^{\nu}\fb} say, in place of 
the total baryon number current {\bb\nn_{\rm b}^{\nu}\fb} above, a 
precisely analogous relation applies in the context of electron 
conductivity in ordinary solid state physics.)

We can now proceed to convert the dynamical energy formula to the generic 
form 
{\be \UU_{\rm dyn}=\frac{_1}{^2}\gamma_{ij}\big(\rrho^{\rm ff}\,
v_{\rm f}^{\,i}\,v_{\rmf}^{\,j}+2\rrho^{\rmfc}\,v_{\rmf}^{\,i}\,
v_{\rmc}^{\,j}+\rrho^{\rm cc}\,v_{\rmc}^{\,i}
\,v_{\rmc}^{\,j}\big)\, ,\label{66}\fe}
in which the required mass density matrix components can be read out as
{\be  \rrho^{\rm ff}={\calK}^{\rm ff}\,\nn_{\rmf}{^2}\, ,\hskip 1 cm
\rrho^{\rmfc}={\calK}^{\rmfc}\,\nn_{\rmf}\nn_{\rmc}\, ,\hskip 1 cm
\rrho^{\rm cc}={\calK}^{\rm cc}\,\nn_{\rmc}{^2}\, .\label{65}\fe}

This expression \rf{66}\fr is to be compared with the alternative 
arrangement whereby the decomposition
{\be \rrho=\rrho_{\rmf}+\rrho_{\rmc}\, ,\label{70}\fe}
of the total mass density {\bb  \rrho = \mm\, \nn_{\rm b}\, ,\fb}
determines a corresponding decomposition
{\be \UU_{\rm dyn}= \tilde\UU_{\rm kin}+\tilde\UU_{\rm ent} 
\, ,\label{67}\fe}
in which the first term is a kinetic energy contribution of the
standard form
{\be \tilde\UU_{\rm kin}=\frac{_1}{^2}\big(\rrho_{\rmf}\,v_{\rmf}{^2}+
\rrho_{\rmc}\,v_{\rmc}{^2}\big)\, ,\label{69}\fe}
while the second term {\bb \tilde \UU_{\rm ent}\fb} in \rf{67}\fr is a frame
independent ``entrainment'' contribution  that can depend only on the
magnitude {\bb \bar v_{\rmfc}\fb} of the relative velocity  \rf{47}\fr and
must therefore be given in terms of some ``entrainment'' mass density
{\bb \bar\rrho_{\rmfc}\fb} by
{\be \tilde\UU_{\rm ent}=\frac{_1}{^2}\bar\rrho_{\rmfc}\,\bar v_{\rmfc}{^2}
\, .\label{68}\fe}
Substituting \rf{68}\fr and \rf{69}\fr in \rf{67}\fr, and comparing
the result with the preceding expression \rf{66}\fr for
{\bb \UU_{\rm dyn}\fb}, we see that the density coefficients in the latter
will be given by
{\be \rrho^{\rm ff}=\rrho_{\rmf}+\bar\rrho_{\rmfc} \, ,\hskip 1 cm
\rrho^{\rmfc}= -\bar\rrho_{\rmfc} \, ,\hskip 1 cm
\rrho^{\rm cc}=\rrho_{\rmc}+\bar\rrho_{\rmfc} \, .\label{73}\fe}
It is to be observed that the values of all the density coefficients
involved will of course depend on the choice of the chemical gauge parameter 
{\bb \Alf_\rmc\fb} that is used for the specification
{\be \rrho_{\rmf}= \mm\, \nn_{\rmf}=\mm(\nn_\rmb-\Alf_\rmc\,\nn_\rmp)
\, \hskip 1 cm
\rrho_{\rmc}= \mm \,\nn_{\rmc}=\mm \Alf_\rmc\,\nn_\rmp\, .\label{72}\fe}
of the relevant ``bare'' mass densities in the decomposition \rf{70}\fr.

The foregoing relations can be translated into the terminology of
effective masses by introducing the values {\bb \mm^{\rmf}_\star\fb} and
{\bb \mm^{\rmc}_\star\fb} according to the specifications
{\be \rrho^{\rm ff}= \nn_{\rmf}\, \mm^{\rmf}_\star\, ,\hskip 1 cm
\rrho^{\rm cc}= \nn_{\rmc}\, \mm^{\rmc}_\star\, ,\label{74}\fe}
so that one obtains
{\be \gamma^{ij}\mmu^\rmf_{\, j}=\mm^\rmf_\star(\vv_\rmf^{\,i}-
\vv_\rmc^{\,i})+\mm \vv_\rmc^{\,i}\, ,\hskip 1 cm
\gamma^{ij}\mmu^\rmc_{\, j}=\mm^\rmc_\star(\vv_\rmc^{\,i}-
\vv_\rmf^{\,i})+\mm \vv_\rmf^{\,i}\, .\fe}
The respective deviations, {\bb \mm^\rmf_{\ \rmc}\fb} and
{\bb \mm^\rmc_{\ \rmf}\fb } say, of these effective masses from the 
ordinary baryonic mass {\bb \mm\fb} will be given in terms of the 
entrainment density by the familiar expressions \cite{Prix01}
{\be\mm^\rmf_{\ \rmc}= \mm_\star^{\rmf}-\mm=\frac{\bar\rrho_{\rmfc}}
{ \nn_{\rmf}} \, ,\hskip 1 cm  \mm^\rmc_{\ \rmf}=\mm_\star^{\rmc}-
\mm=\frac{\bar\rrho_{\rmfc}}{\nn_{\rmc}}  \, .\label{75}\fe}
It can be seen from \rf{75}\fr that \emph{both} effective masses are 
either larger or smaller than {\bb\mm ,\fb} but one effective mass
cannot be larger than {\bb\mm\fb} and the other smaller. In particular
whenever one of the effective masses coincides with the ordinary baryon 
mass, this entails that the other one is also equal to the ordinary mass.

It is important to distinguish the true (current) 3-velocities used above
from the pseudo-velocities that are referred to in many older and even
recent discussions~\cite{Lindblom00} as ``superfluid velocities'', and
that are constructed by dividing the corresponding 3-momentum by the
relevant mass parameter. The example that arises in the present context is
that of the so called ``superfluid neutron velocity''
{\bb V^{_{\rm S}}_{\, i}\fb} which is obtained
by setting {\be \mmu^{\rmf}_{\, i}=\mm\, V^{_{\rm S}}_{\, i}\, .\fe}
The corresponding ``normal velocity'' {\bb \vv_{_{\rm N}}^{\, i}\fb} is
not a pseudo velocity but a true velocity in the sense of being the space
projected part of a physically well defined 4-velocity, namely that of the
crust reference frame as given by {\bb \uu_\rmc^{\,\nu} ,\fb} so that we
can make the identification  {\bb \vv_{_{\rm N} }^{\,i}=\vv_\rmc^{\,i} .\fb}
This leads to a corresponding density decomposition
{\be \rrho=\rrho_{_{\rm S}}+\rrho_{_{\rm N}}\, ,\fe}
(which unlike \rf{70}\fr is not affected by the choice of the gauge
parameter {\bb\Alf_\rmc\fb} as will be shown later) in terms of which the 
dynamical energy will be expressible simply as
{\be \UU_{\rm dyn}=\frac{_1}{^2}\big(\rrho_{_{\rm S}}\, V^{_{\rm S}2}+
\rrho_{_{\rm N}}\,v_{_{\rm N}}{^2}\big)\, ,\fe}
with the so called ``superfluid density''  {\bb \rrho_{_{\rm S}}\fb}
and the so called ``normal density''  {\bb \rrho_{_{\rm N}}\fb} given
in the notation of \rf{75}\fr by
{\be  \rrho_{_{\rm S}}=\rrho_{\rm f}\,\mm/\mm^{\rmf}_\star
\, ,\hskip 1 cm \rrho_{_{\rm N}}= \rrho_{\rm c}+
\rrho_{\rm f}\,\mm^{\rmf}_{\ \rmc}\,/\mm^{\rmf}_\star
\, .\label{superdens}\fe}

\section{Underlying physical theory of neutron conduction}

\subsection{Microscopic analysis in the crust frame}

As it is not practical to carry out direct experimental measurements
for bulk matter at neutron star densities, the evaluation of the
relevant equation of state functions requires a theoretical analysis 
based on an underlying physical model. In particular, to provide the
quantitative estimates~\cite{Chamel04} needed for obtaining the 
entrainment density, or the corresponding neutronic effective mass 
in the ionic crust layers above the neutron drip density threshold, it
is necessary to use a microscopic model of the kind whose 
essential elements were presented in our preceding work~\cite{CCHI}
and that we have recently developed in a follow
up~\cite{CCHIII} allowing for the effect of BCS pairing.

The approach we use provides a typical continuum fluid description as
formulated in terms of mesoscopically homogeneous configurations that are 
obtainable from a corresponding microscopic theory by the minimisation 
subject to relevant constraints of the energy density {\bb\UU\fb} that is 
defined in terms of a quantum system in a unit volume sample box (subject 
to periodic boundary conditions) as the averaged expectation value, which
we indicate by angle brackets,  of the relevant total Hamiltonian operator
{\bb\hHH\fb} say, i.e. {\bb \UU=\langle\hHH\rangle\fb}. The analysis will be
carried out using the crust rest frame, in which the heavy ionic nuclei 
can be treated as classical particles at fixed positions, to which all
the protons and a subset of ``confined'' neutrons are considered to be 
bound. For simplicity a quantum description is applied only to ``free'' 
(unconfined) neutrons (whose role is analoguous to that of ``conduction'' 
electrons in ordinary solid state theory) whose Bragg scattering by the  
potential wells associated with the nuclear clusters gives rise to strong 
entrainment effects.

In a strictly static configuration the only relevant constraint is the 
preservation of the number density {\bb \nn\fb} of such unconfined 
neutrons, which is definable as the corresponding average 
{\bb \nn=\langle\hnn\rangle\fb} of a conserved particle number operator 
{\bb\hnn\fb} say. The minimisation of  {\bb \UU\fb} subject to such a 
constraint is equivalent to the absolute minimisation of a combination of 
the form
{\be \UU^\prime=\UU-\mmu\,\nn\, ,\fe}
in which {\bb\mmu\fb} is a Lagrange multiplier that will be interpretable as
the relevant chemical potential. For any given value of {\bb\nn\fb} the
corresponding constrained minimum state can be expected on symmetry grounds
to be static, in the sense of containing no relatively moving currents. To
obtain the relatively conducting configurations in which we are interested
here, we need to impose the further constraint  of preservation of the
relevant current components {\bb \nn^i\fb} (with respect to the rest frame
of the sample box, \textit{i.e.} in the crust frame) as defined by the
corresponding averages of the relevant quantum operators
{\bb \nn^i=\langle\hnn^i\rangle\fb} say. The problem will thus be that of 
minimising a combination of the form
{\be \UU^\prime_{\{\pp\}}=\UU^\prime-\pp_i\,\nn^i\label{8}\fe}
in which the quantities {\bb\pp_i\fb} are Lagrange multipliers that will
be seen to be interpretable as representing effective momentum per particle,
so that an infinitesimal variation in the neighbourhood of such a
conditionally minimised energy density will be given by
{\be \delta\UU=\mmu\,\delta\nn+\pp_i\delta\nn^i\, .\label{9}\fe}
This means that the multipliers {\bb\mmu\fb} and {\bb\pp_i\fb} can be
construed as partial derivatives with respect to number density and current
respectively.

Subject to the realistic assumption that the current is small, the energy
density will be close to the static internal energy value
{\bb \UU_{\rm ins}\fb} say (minimising {\bb\UU^\prime\fb}) in the state of
 zero current characterised by a given value of the number density 
{\bb \nn\fb} and the ionic charge number {\bb \ZZ\fb}. It will therefore
be expressible by an expansion that will be given to second order in the 
current density {\bb \nn^i\fb} by a decomposition of the form
{\be \UU= \UU_{\rm ins}+\UU_{\rm dyn}\, ,\label{22}\fe}
in which -- for a given value of the charge number {\bb \ZZ\fb}
and the confined particle number density {\bb \nn_{\rmc}\fb}  which, in the
present case are considered as given classical variables in the
microscopic model -- the static internal energy density
{\bb \UU_{\rm ins}\fb} will depend just on the relevant particle number 
density {\bb \nn\fb}, so that it will be expressible as
{\be \UU_{\rm ins}= \UU_{\rm ins}\{\ZZ,\nn_{\rmc},\nn\}\, ,\label{22a}\fe}
while the dynamical contribution is given in terms of a positive
definite matrix with tensor components {\bb {\calK}^{ij}\fb} that also
depend just on the same three scalar variables by an expression having
the homogeneous quadratic form
{\be \UU_{\rm dyn}=\frac{_1}{^2}{\calK}^{-1}_{\,ij}\nn^i \nn^j
\, . \label{22b}\fe}
By differentiating this expression with respect to the current components,
it can be seen that in terms of the momentum components appearing as
partial derivative coefficients in the variation formula \rf{9}\fr, the
current  will be given by the formula
{\be \nn^i=\calK^{ij}\pp_j\, .\label{18}\fe}

In practice {\bb\nn\fb} and {\bb \nn_{\rmc}\fb} can be expected to remain
close to values that are determined as functions of {\bb\ZZ\fb} and the 
total baryon number density {\bb \nn_{\rm b}\fb} by a condition of 
chemical equilibrium whereby {\bb \UU_{\rm ins}\fb} is minimised, with 
value {\bb \UU_{\rm eq}\fb} say, for the given values of {\bb\ZZ\fb} and
{\bb \nn_{\rm b}\fb}, where {\bb\nn\fb} will have a value, 
{\bb\nn_{\rm eq}\fb} say, that is determined as a function of {\bb\ZZ\fb}
 and {\bb \nn_{\rm b}\fb} and hence of the corresponding value of
{\bb \nn_{\rmc}\fb}. Therefore rather than considering the coefficients
{\bb {\calK}^{ij}\fb} to be functions of the three independent variables
{\bb\ZZ ,\fb}{\bb \nn_{\rmc}\fb} and {\bb\nn ,\fb} it should be an 
adequate approximation to consider them to be functions just of the pair 
of variables {\bb\ZZ\fb} and {\bb \nn_{\rmc}\fb} that determine the
classical background field in the model. This allows  us to deduce 
from \rf{9}\fr that the partial derivative with respect to  {\bb\nn\fb} of
the  static internal energy density function \rf{22a}\fr will be given
exactly by
{\be \frac{\partial  \UU_{\rm ins}}{\partial \nn}=\mmu\, .\label{18a}\fe}
To accuracy of linear order in deviations from chemical equilibrium it
will therefore suffice to take the function \rf{22a}\fr to have the 
simple form
{\be \UU_{\rm ins}= \UU_{\rm eq}+\mmu(\nn-\nn_{\rm eq})\label{18b}\fe}
in which  {\bb \UU_{\rm eq}\fb}, {\bb\nn_{\rm eq}\fb} and {\bb \mmu\fb}
itself are all functions just of the pair of variables 
{\bb \nn_{\rmc}\fb} and {\bb\ZZ\fb} that characterise the mean density 
and degree of clustering of the underlying distribution of protons.

The mobility tensor {\bb {\calK}^{ij}\fb} defined by the formula 
\rf{18}\fr might be anisotropic in a solid \cite{CC06} of perfectly 
regular crystalline type, but one would expect it to be given by an 
expression of the isotropic form
{\be {\calK}^{ij}={\calK}\gamma^{ij}\, ,\label{25}\fe}
in which the scalar coefficient will evidently be given by
{\be {\calK}=\frac{_1}{^3}\gamma_{ij}{\calK}^{ij}\, ,\label{26}\fe}
not only for a medium that is a liquid (as will be the case in a neutron
star crust when the star is very young) and for the case of a solid having
a glasslike or disordered crystalline structure on a macroscopic scale
(as is likely to be the case in a realistic description of a neutron star
crust) but even for a perfectly regular crystal lattice provided it is
of a cubic type (such as will be energetically favoured throughout the
neutron star crust except near the base where rod or slab like structures
may be preferred).

In terms of the mobility scalar introduced in this way, the relation
between current and momentum (in the crust frame) will be given by
{\be \nn^i={\calK}\gamma^{ij}\pp_j\, ,\label{27}\fe}
and the corresponding final result for the dynamical energy contribution
in \rf{22b}\fr will be given by
{\be \UU_{\rm dyn}= \frac{_1}{^2} {\calK}\gamma^{ij}\pp_i\pp_j
= \frac{_1}{^2} {\calK}^{-1}\gamma_{ij}\, \nn^i \nn^j\, .\label{28}\fe}

\subsection{Matching with the macroscopic description}

In order to match the microscopic quantum mechanical description in
the previous subsection to the macroscopic fluid description summarised
before, an obvious first step is to take the convention that
the fraction of the neutrons that is considered to be bound
to the ionic nuclei is such that we can identify the unbound neutron 
nunber density {\bb\nn\fb} with the free neutron number density 
{\bb\nn^{\rmf}\fb} introduced above.

Having thus set
{\be \nn^\rmf=\nn\fe}
we go on to observe that in the crust frame characterised by
{\bb  \vv_{\rmc}^{\,i}=0\fb} in which the preceding microscopic analysis
was carried out the free neutron current will just be the same as
the (gauge independent) conduction current \rf{48}\fr so we shall be able 
to make the further identification
{\bb \nn^i=\nn_{\rmf}^{\,i}\, .\fb} 

It can be seen that with respect to this particular frame the kinetic
energy will simply be given by 
{\be \tilde\UU_{\rm kin}=\frac{_1}{^2}\,\mm\, \nn_\rmf\, \gamma_{ij}\,
\vv_\rmf^{\,i}\,\vv_\rmf^{\,j}=\frac{_1}{^2}\,\frac{\mm}{\nn_\rmf}\,
\gamma_{ij}\,\nn^i\nn^j\, .\fe}
In order to evaluate the (frame independent) entrainment energy that is 
definable, in accordance with \rf{67}\fr as
{\be \tilde \UU_{\rm ent}=\UU_{\rm dyn}-\tilde \UU_{\rm kin}\, ,\fe}
it  now suffices to substitute the formula \rf{22b}\fr for the
dynamical part, which provides the required result in the form
{\be \tilde\UU_{\rm ent}=\frac{_1}{^2}\left(\calK^{-1}_{\, ij}-
\frac{\mm}{\nn_\rmf}\gamma_{ij}\right)\nn^i\nn^j
\, ,\label{27B}\fe}
which is noteworthy for remaining valid for application to the elastic 
solid models \cite{CC06} that will ultimately be needed, in which the
mobility tensor {\bb\calK^{ij}\fb} need not always be isotropic.

In the fluid case characterised by \rf{25}\fr with which we are
concerned here, the formula \rf{27B}\fr reduces to
{\be \tilde\UU_{\rm ent}=\frac{_1}{^2}\left(\frac{1}{\calK}-
\frac{\mm}{\nn_\rmf}\right)\gamma_{ij}\,\nn^i\nn^j
\, .\label{27C}\fe}
By comparison with \rf{68}\fr the entrainment density can be 
read out from this as
{\be \bar\rrho_{\rmfc}=\nn_{\rmf}{^2}/ \calK -\rrho_{\rmf}
\, ,\label{80}\fe}
which by \rf{73}\fr a implies
{\be \rrho^{\rm ff} =\nn_\rmf^{\,2}/{\calK}\, .\fe}
This quantity depends on the choice of chemical basis, but we 
can use \rf{65}\fr to obtain the corresponding coefficient
{\be {\calK}^{\rm ff} ={\calK}^{-1}\, ,\label{76}\fe}
which has the important property of being gauge invariant in the 
sense that it is independent of the choice of the gauge parameter 
{\bb\Alf_\rmc\fb} that has been used in \rf{chemtrans}\fr for 
specifying the number density {\bb\nn_\rmf\fb} of neutrons that are 
counted as ``free''. The gauge independence of the tensor 
{\bb \calK ^{-1}_{\,ij}\fb} and thus of the scalar {\bb \calK \fb} is 
an evident consequence of the gauge independence of the relative 
current vector \rf{50}\fr in terms of which it was defined 
by \rf{22b}\fr.

We can go on to rewrite the formulae \rf{superdens}\fr for the so called
``superfluid'' and ``normal'' density contributions in the manifestly
gauge independent form
{\be  \rrho_{_{\rm S}}=\rrho-\rrho_{_{\rm N}}=\mm^2\calK\, .\fe}
This shows that the ``superfluid'' density contribution  
 {\bb \rrho_{_{\rm S}}\fb} will tend to zero when the mobility
coefficient {\bb\calK \fb} tends to zero, which is what occurs at the 
``neutron drip'' transition (whose role is thus analogous to that of the 
``lambda point'' transition in ordinary liquid helium).

The actual evaluation of the required mobility coefficient {\bb\calK\fb} 
as a function of the relevant densities has been initiated in the 
preceding work~\cite{CCHI} which describes the way in which a simple 
nuclear physical treatment can be applied most easily to the rod or plate 
type configurations that are likely to be relevant at the base of the 
crust~\cite{PR95, Hae01}, while more recent work using more elaborate numerical
analysis has extended the range of this treatment to include three 
dimensional cubic configurations as well \cite{Chamel04}. The quantitative
results obtained so far are of an approximate provisional nature, and  
much further work will be needed to improve their precision and
reliability. A first step towards such refinement has been taken in our
recent examination ~\cite{CCHIII} of the -- apparently only moderate --
adjustment of the mobility tensor that is needed to take account of the
pairing effect responsible for the BCS type superfluidity of the neutrons
that is predicted to occur in the low to moderate temperature range that 
is relevant in pulsars.

When the appropriate value of {\bb\calK\fb} has been obtained, the 
complete determination of the model still requires the specification of 
the (frame independent) internal static contribution \rf{58}\fr. This can 
evidently be obtained just by adopting the prescription \rf{18b}\fr, 
which  enables us to make the simple identification
{\bb \mmu^{\rmf}_{_0}\rightarrow -\mmu\, \fb} in the zero
current limit.

\section{Restoration of frame covariance} 

\subsection{Chemically invariant action formula}

In the crust rest frame used in the previous section the ``free''
neutron momentum covector was just  the quantity
{\bb\pp_i\fb} that is given by \rf{27}\fr, so with respect to a generic
frame with non vanishing 3 velocity {\bb \vv_\rmc^{\,i}\fb} this 
quantity will be given by the formula
{\be \mmu^{\rmf}_{\,i}=\mm\gamma_{ij}
\vv_{\rmc}^{\,j} +\pp_i \, ,\label{79}\fe}
in which it is to be observed that both terms are  independent of the
parameter $\Alf_\rmc$ that specifies the chemical gauge, so
that as remarked in Subsection \ref{chemin} their is no
ambiguity in the application of the superfluidity condition to
the effect that the momentum covector \rf{79}\fr should be irrotational.

This contrasts with the chemical gauge dependent status of the corresponding
``crust'' momentum  covector which will be expressible
in terms of the (chemically gauge invariant) relative
conduction current {\bb \nn^i \fb} \rf{48}\fr by
{\be \mmu^{\rmc}_{\,i}=\mm\gamma_{ij}
\vv_{\rmc}^{\,j} +\frac{\mm}{\nn_{\rmc}}\gamma_{ij} \nn^j  
-\frac{ \nn_{\rmf}}{\nn_{\rmc} } \pp_i \, .\label{79bis}\fe}
in which the relative conduction current {\bb \nn^{\, i}\fb} and the 
corresponding momentum contribution {\bb\pp_i\fb} are gauge invariant, 
but the densities {\bb\nn_{\rmf}\fb} and {\bb\nn_{\rmc}\fb} are not.
 
Similar remarks apply to the complete dynamical action density
 \rf{dynamic}\fr which can be given by the formula
{\be\UU_{\rm dyn}= \frac{_1}{^2}\rrho\,\vv_\rmc^{\,2}+
\calK\left(\mm\,\vv_\rmc^{\,i}\pp_i+ \frac{_1}{^2}\pp^2 \right)\, ,\fe}
in which each separate term is chemically invariant, in contrast with
the status of the separate kinetic and entrainment contributions
in the decompositions \rf{67}\fr, and in the corresponding decomposition
{\be\Lamb_{\rm mat}= \tilde \Lamb_{\rm int}+\tilde\Lamb_{\rm kin}
\label{Lmat}\fe}
of the material action contribution in which the kinetic part is simply 
given by {\bb \tilde\Lamb_{\rm kin} = \tilde\UU_{\rm kin} \fb} and the 
internal contribution will take the form
{\be\tilde \Lamb_{\rm int}=\tilde \UU_{\rm ent}-\tilde\UU_{\rm ins}
\, .\label{Lint}\fe}
This quantity needs to be distinguished from the
corresponding internal energy density defined by
{\be\tilde\UU_{\rm int}=\UU_{\rm mat}-\tilde\UU_{\rm kin}\, ,\fe}
which will be given by
{\be\tilde\UU_{\rm int}=\tilde\UU_{\rm ent}+\tilde\UU_{\rm ins} \, .
\label{Uint}\fe}
In particular it can be seen from \rf{Lint}\fr and \rf{Uint}\fr that the 
internal contribution {\bb \tilde\Lamb_{\rm int} \fb}  to the Lagrangian 
density is not simply given by the negative of the internal energy density
 {\bb \tilde\UU_{\rm int} \fb} (as stated by
 Prix \textit{et al.} \cite{PrixComAnd, Prix01, Rieutord02, Prix02})
 but takes the form
{\be\tilde \Lamb_{\rm int}=-\tilde \UU_{\rm int}+2\tilde\UU_{\rm ent} 
\, . \fe}
The tilde is used here as a reminder that the contributions thus
designated have the disadvantage of being chemically gauge dependent.
However the terms in \rf{Lint}\fr  have the compensating advantage of
Galilean frame invariance. It is therefore convenient and customary to use
these entities as a starting point for the specification of particular
models, as they contain just the minimum information, in the form of the
particular equations of state,  that are needed for this purpose, namely
expressions as a function of the relevant scalar densities
{\bb\nn_{\rmf}\fb}, {\bb\nn_{\rmc}\fb} and if necessary also
{\bb\nn_{\rmI}\fb}, for the static internal energy
{\bb\UU_{\rm ins}\fb} and for the mobility coefficient {\bb\calK\fb}
or equivalently, via \rf{80}\fr, of the entrainment density
{\bb\bar\rrho_{\rmfc}\fb} for the entrainment part {\bb \tilde\UU_{\rm ent}
\, . \fb}
In terms of the relative velocity defined by \rf{47}\fr the generic variation 
of the frame invariant internal action density will evidently be expressible 
in the form
{\be \delta \tilde\Lamb_{\rm int}=\bar\rrho_{\rmfc}\,\bar v_{\rmfc\, i}\,
\delta \bar v_{\rmfc}^{\, i}-\tilde\cchi^{\rmf}\delta\nn_\rmf-
\tilde\cchi^{\rmc}\delta\nn_\rmc-\tilde\cchi^{\rmI}\delta\nn_\rmI
\, .\label{entvar}\fe}
The frame independent chemical potentials {\bb\tilde\cchi^{\rmf}\fb}
{\bb\tilde\cchi^{\rmc}\fb}, {\bb\tilde\cchi^{\rmI}\fb} defined in this
way determine corresponding potentials
{\be\tilde\mmu^{\rmf}= \tilde\cchi^{\rmf}-\frac{_1}{^2}m\, v_\rmf^{\, 2}
\, ,\hskip 1 cm\tilde\mmu^{\rmc}= \tilde\cchi^{\rmc}-\frac{_1}{^2}m\,
v_\rmc^{\, 2}\, ,\hskip 1 cm\tilde\mmu^{\rmI}= \tilde\cchi^{\rmI}
\, ,\fe}
(of which the first two are Galilean frame dependent) in terms
of which the variation of the complete (chemical gauge invariant)
material action density \rf{Lmat}\fr will take the form
{\be \delta\tilde\Lamb_{\rm mat}=\nn_\rmf\mmu^\rmf_{\, i}
\delta v_{\rmf}^{\ i}+\nn_\rmc\mmu^\rmc_{\, i}
\delta v_{\rmc}^{\ i}-\tilde\mmu^{\rmf}\delta\nn_\rmf-\tilde\mmu^{\rmc}
\delta\nn_\rmc-\tilde\mmu^{\rmI}\delta\nn_\rmI\, .\fe}
To match this with the preceding expression \rf{deLamb}\fr for
the material action variation, it can be seen that we need to make the 
identifications
{\be \tilde\mmu^{\rmf}=-\uu_\rmf^{\,\nu}\mmu^\rmf_{\,\nu}\, ,\hskip 1 cm
\tilde\mmu^{\rmc}=-\uu_\rmc^{\,\nu}\mmu^\rmc_{\,\nu}\, ,\fe}
(whereas to relate this to the terminology of Prix \textit{et al.} 
\cite{Prix01} one would need to make the translations {
\bb \tilde\cchi^{\rmf} \rightarrow \mu^{\rmf} ,\fb} 
{\bb \tilde\cchi^{\rmc}\rightarrow \mu^{\rmc}\fb} and 
{\bb \tilde\Lamb_{\rm int}\rightarrow -{\cal E} \fb}  ).

To match this with an alternative treatment of a kind suitable for
generalisation \cite{CC06} to allow for solid elasticity, in which the 
confined constituent will have a privileged role, the  relative velocity 
{\bb \bar v_{\rmfc}^{\, i} \fb} needs to be replaced in the variation 
\rf{entvar}\fr by the relative current {\bb\nn^i \fb} given by \rf{48}\fr 
which has the advantage of being chemically invariant. 
In the notation of \rf{75}\fr one thus obtains
{\be \delta \tilde\Lamb_{\rm int}=\frac{\mm^\rmf_{\ \rmc}}{\nn_\rmf}
\,\gamma_{ij}\,\nn^i \,\delta\nn^j-\cchi^{\rmf}\delta\nn_\rmf-
\cchi^{\rmc}\delta\nn_\rmc-\cchi^{\rmI}\delta\nn_\rmI\, ,\fe}
with
{\be \cchi^{\rmf}=\tilde\cchi^{\rmf}+\mm^{\rmf}_{\ \rmc}\,
\bar v_{\rmfc}^{\,2} \, ,\hskip 1 cm\cchi^{\rmc}=\tilde\cchi^{\rmc}
\, ,\hskip 1 cm \cchi^{\rmI}=\tilde\cchi^{\rmI}\, .\fe}

\subsection{Effective mass relations}

 Subject to the understanding that number density {\bb \nn\fb} in
the preceeding  work ~\cite{CCHI, Chamel04} is to be interpreted as the
``free'' number density {\bb \nn_{\rmf}\fb} in the present terminology,
the corresponding effective mass {\bb \mm_\star\fb} will be identifiable
with the effective mass per free particle as denoted here by
{\bb \mm^{\rmf}_\star\fb}, which means that it will be given according to
\rf{76}\fr  by
{\be \mm^{\rmf}_\star=\nn_{\rmf}/\calK \, .\label{78}\fe}
A particularly noteworthy and previously unexpected conclusion that has
emerged from our preliminary investigations~\cite{CCHI, Chamel04} is
this effective mass can be expected to become very large (by a factor of
several hundred percent)  compared with the ordinary neutron mass
{\bb \mm\fb} in the middle layers of the outer crust, even when it is only
the neutrons outside the nuclei that are counted as free. It would of
course be even larger with respect to the simple ``comprehensive'' gauge
given by {\bb\Alf_\rmc=1\fb} for which all the neutrons are counted as
 ``free''.

Before concluding, it needs to be pointed out that whereas the definitions
used here for the effective masses are  equivalent to setting
{\be {\mm_\star^{\rmf}}={\calK}^{\rm ff}{ \nn_{\rmf}} 
\, ,\hskip 1 cm{\mm_\star^{\rmc}}={\calK}^{\rm cc}
{\nn_{\rmc}}  \, ,\label{81}\fe}
there is an alternative definition that is commonly used in the published 
literature on neutron star matter~\cite{Mendell91}. This alternative
definition can be formulated in the present terminology by setting
{\be \mm^{\rmf}_\sharp  ={\nn_{\rmf}}/{\calK_{\rm ff}} \, ,\hskip 1 cm
\mm^{\rmc}_\sharp  ={\nn_{\rmc}}/{\calK_{\rm cc}} 
 \,  .\label{82}\fe}
The effective masses of this second kind will be given in terms of those
of the first kind by
{\be \mm^{\rmf}_\sharp-\mm=\frac{\mm}{\mm_\star^{\rmc}}
\big(\mm^{\rmf}_\star-\mm\big)\, ,\hskip 1 cm  \mm^{\rmc}_\sharp-\mm=
\frac{\mm}{\mm_\star^{\rmf}} \big(\mm_\star^{\rmc}-\mm\big)\, .\fe}
These effective masses of the second kind are (like the first kind) either
both larger or both smaller than the ordinary mass.

To complete the specification of the two fluid description, the coefficient
{\bb {\calK}^{\rmfc}\fb} giving rise to entrainment can be seen to be given
by either of the equivalent formulae
{\be  {\calK}^{\rmfc}\nn_\rmc=-\mm^\rmf_{\ \rmc}=
\mm - \mm_\star^{\rmf}\, ,\hskip 1 cm {\calK}^{\rmfc}\nn_\rmf
=-\mm^\rmc_{\ \rmf}= \mm - \mm_\star^{\rmc}\, , \label{84}\fe}
Thus {\bb {\calK}^{\rmfc}\fb} will vanish, as expected, whenever the 
effective masses are equal to the ordinary (protonic) baryon mass 
{\bb\mm\fb}. (As previously remarked, both effective masses are equal to 
{\bb \mm\fb}  whenever one of them is  equal to {\bb\mm\fb}).  The 
alternative description in terms of the effective masses of the second kind 
leads to the expressions
{\be {\calK}_{\rmfc} =\frac{\nn_{\rmf}}{\mm}\,
(1-\frac{\mm}{\mm^{\rmf}_\sharp})  =\frac{ \nn_{\rmc}}{\mm}\,
 (1-\frac{\mm}{\mm^{\rmc}_\sharp}) \label{85} .\fe}
The conditions \rf{64bis}\fr can thereby be restated in terms of effective
masses in the equivalent forms
{\be  \frac{\mm_\star^{\rmc}}{\mm} > \frac{\nn_{\rmc}}{\nn_{\rmb}}\, , 
\hskip 1 cm \frac{\mm_\star^{\rmf}}{\mm} > \frac{\nn_{\rmf}}{\nn_{\rmb}}
\label{86} \, , \fe}
while the effective masses of the second kind must obey the 
equivalent inequalities
{\be \frac{\mm^{\rmc}_\sharp}{\mm} < \frac{\nn_{\rmb}}{\nn_{\rmc}}\, ,
\hskip 1 cm \frac{\mm^{\rmf}_\sharp}{\mm} < \frac{\nn_{\rmb}}{\nn_{\rmf}}
\label{87}\, . \fe}

\subsection{Comprehensive treatment of inner crust and (outer) core}

The approach developed above has been particularly designed for
the treatment of the inner crust layers above the ``neutron drip''
density threshold but below the transition density at which the ionic 
clusters merge. It is desirable to relate this approach to that of 
related work\cite{PrixComAnd, Prix01, Prix02} designed more specifically
for treatment of the homogeneous (unclustered) nuclear matter
of the core region, or to be more precise the outer core region
(since the most massive neutron stars may also contain an inner core
consisting of hyperons or quark matter that would need a relativistic 
treatment of kind that is beyond the scope of the present discussion).

Instead of using the more general kind of chemical basis that is
useful for discussion of the ``neutron drip'' transition, for the 
purpose of matching to the formalism that is usually employed for work 
on the outer core it is convenient to work exclusively in the 
``comprehensive'' gauge characterised by setting {\bb\Alf_\rmc =1\fb} 
in \rf{chemtrans}\fr, which simply means identifying the ``free'' 
neutron current {\bb \nn_\rmf^{\,\nu}\fb} with the entire neutron 
current  {\bb \nn_\rmn^{\,\nu}\fb} and identifying the ``confined'' 
baryon current   {\bb \nn_\rmc^{\,\nu}\fb} with the proton current 
{\bb \nn_\rmp^{\,\nu}\fb}. This allows us to obtain a treatment 
combining the representation of the crust and the core in a single 
``comprehensive'' model in which the relevant ``comprehensive'' 
effective mass, 
{\be \mm_\star^\rmn =\nn_\rmn/\calK \, ,\label{88} \fe} 
(of the first kind) for the neutrons has the advantage of having a 
clear physical definition, though --  as the price for this -- it has
the awkward property of diverging to infinity as the density decreases 
to that of the ``neutron drip'' threshold.

\begin{figure}
\centering
\epsfig{figure=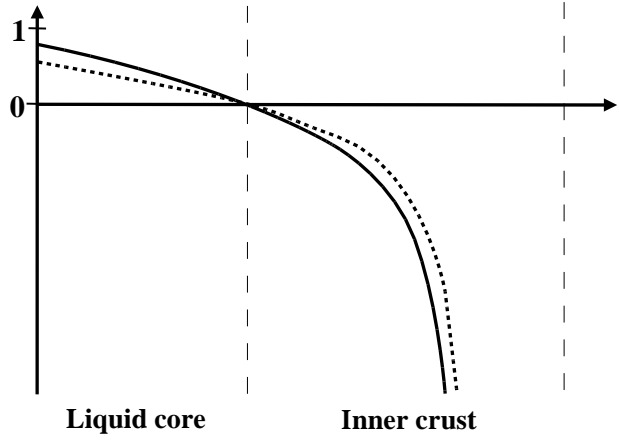, height=6 cm}
\caption{Rough sketch of expected radial dependence of
dimensionless ``comprehensive'' entrainment coefficients
{\bb \varepsilon_\rmn \fb} (dashed line)
and {\bb \varepsilon_\rmp \fb} (solid line) in a neutron star.}
\label{eps_np}
\end{figure}

\begin{figure}
\centering
\epsfig{figure=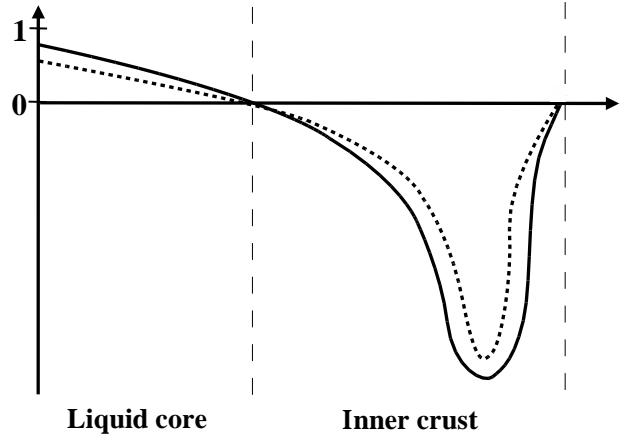, height=6 cm}
\caption{Rough sketch of expected radial dependence of
dimensionless ``operational'' entrainment coefficients
{\bb \varepsilon_\rmf \fb} (dashed line)
and {\bb \varepsilon_\rmc \fb} (solid line) in a neutron star.}
\label{eps_fc}
\end{figure}

The link with work on the core regime can be made more explicit by 
using notation of the kind employed by Prix \textit{et al.} 
\cite{PrixComAnd, Prix01, Rieutord02, Prix02} in which the neutron and 
proton 3-momenta are given in terms of the relative 3-velocity 
{\bb {\Deltav}^i = \vv_{\rmp}^i - \vv_{\rmn}^i \fb} by expressions of 
the form 
{\be \mmu^{\rmn}_{\,i}=\gamma_{ij}\mm\big(\vv_{\rmn}^{\,j} +
\varepsilon_\rmn {\Deltav}^j\big)\, ,\hskip 1 cm \mmu^{\rmp}_{\,i}=
\gamma_{ij}\mm\big( \vv_{\rmp}^{\,j} - \varepsilon_\rmp {\Deltav}^j 
\big)\, ,\label{89}\fe}
involving dimensionless parameters {\bb \varepsilon_\rmn \fb} and 
{\bb \varepsilon_\rmp \fb} that can be seen from  \rf{61}\fr and 
\rf{89}\fr to be definable as
{\be \varepsilon_\rmn = (\mm - \mm_\star^{\rmn})/\mm 
\, , \hskip 1cm
\varepsilon_\rmp = (\mm - \mm_\star^{\rmp})/\mm \, . \label{90}\fe}
According to an identity of the same kind as that given by \rf{84}\fr, 
these coefficients {\bb \varepsilon_\rmn \fb} and {\bb \varepsilon_\rmp \fb} 
will be related to each other, and to a density variable {\bb\aalpha\fb}
introduced by Prix \textit{et al.} \cite{Prix01} by
{\be \varepsilon_\rmn \nn_\rmn= \varepsilon_\rmp \nn_\rmp=2\aalpha/\mm\, ,
\hskip 1 cm \aalpha=-\frac{_1}{^2}\,\bar\rrho_{\rm np}
\, .\label{91}\fe} 
This means in particular that for any given layer inside the star, the
coefficients {\bb\varepsilon_\rmp\fb} and {\bb\varepsilon_\rmn\fb} both 
have the same sign, and it also shows that we shall have {\bb 
|\varepsilon_\rmn|< |\varepsilon_\rmp| \fb} since the density of neutrons 
will always exceed the number of protons, increasing from a comparable 
value in the outer layers, where we shall have {\bb \varepsilon_\rmn \approx 
\varepsilon_\rmp ,\fb} to a very much higher value, 
{\bb\nn_\rmn\gg\nn_\rmp\fb} in the neutron rich core where we shall have 
{\bb |\varepsilon_\rmn| \ll |\varepsilon_\rmp| .\fb} In terms of the proton 
fraction {\bb x_\rmp = \nn_\rmp/ \nn_\rmb < 1/2 ,\fb} it  can be seen 
that an inequality of the type \rf{86}\fr is equivalent to either of the 
restrictions
{\be \varepsilon_\rmn < x_\rmp \, ,\hskip 1 cm
\varepsilon_\rmp < 1 - x_\rmp \label{93}\, .\fe} 
which evidently entails that the entrainment coefficients 
{\bb \varepsilon_\rmn \fb} and {\bb \varepsilon_\rmp \fb} must 
both be less than unity.

\begin{figure}
\centering
\epsfig{figure=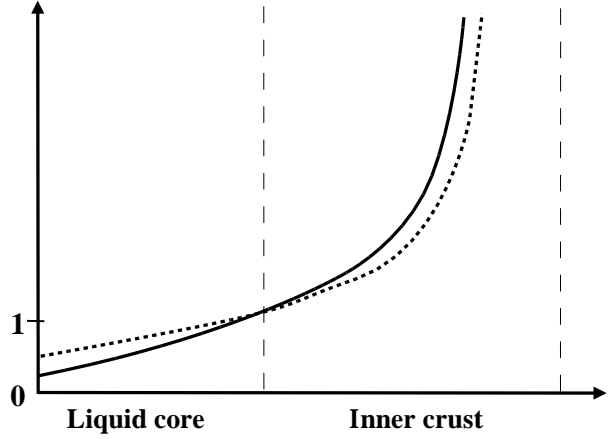, height=6 cm}
\caption{Rough sketch of expected radial dependence of
``comprehensive'' effective mass ratios {\bb \mm_\star^\rmn/\mm \fb} 
(dashed line) and {\bb \mm_\star^{\rmp}/\mm \fb} (solid line) in a 
neutron star.}
\label{effmass_np}
\end{figure}

\begin{figure}
\centering
\epsfig{figure=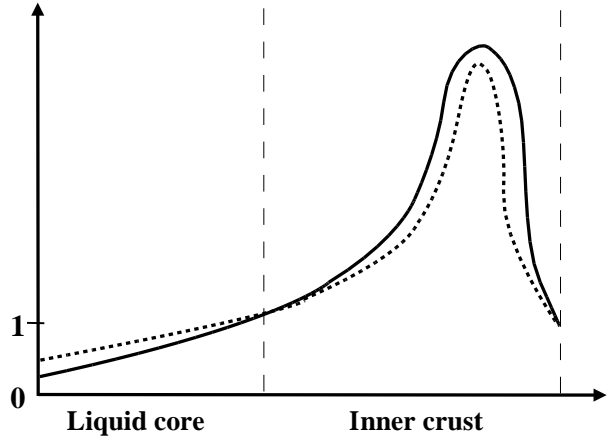, height=6 cm}
\caption{Rough sketch of expected radial dependence of ``operational''
effective mass ratios, {\bb \mm_\star^\rmf/\mm \fb} (dashed line)
and {\bb \mm_\star^\rmc/\mm \fb} (solid line) in a neutron star.}
\label{effmass_fc}
\end{figure}

As the effective mass {\bb \mm_\star^\rmn \fb} (and hence also
{\bb \mm_\star^\rmp \fb}) has been found \cite{BJK96,ComerJoynt03}
to be smaller than the ordinary nucleon mass {\bb \mm \fb}  in the
outer core, it follows that the entrainment coefficients 
{\bb \varepsilon_\rmn \fb} and {\bb \varepsilon_\rmp \fb} will both be 
positive there. This contrasts with situation in the crust, where we expect 
the ``comprehensive'' effective mass  {\bb \mm_\star^\rmn>\mm \fb} to be 
larger compared with {\bb\mm\fb} so that the coefficients  
{\bb \varepsilon_\rmn \fb} and {\bb \varepsilon_\rmp \fb} will both be 
negative, since we have found \cite{CCHI, Chamel04} that the smaller 
effective mass given in a ``realistic'' gauge just for ``conduction''  
neutrons with number density {\bb\nn_\rmf\fb} (defined as those contributing 
significantly to the current) by  {\bb \mm_\star^\rmf = 
\mm_\star^\rmn\nn_\rmf/\nn_\rmn\fb} will already be larger (and in some 
layers much larger) than the ordinary mass {\bb\mm\fb} in the crust. It is 
to be observed that the negative values of {\bb \varepsilon_\rmn \fb} and
{\bb \varepsilon_\rmp \fb} will actually diverge at the ``neutron drip'' 
transition where the ``conduction'' number density {\bb\nn_\rmf\fb} tends 
to zero. These considerations are summarised on figures
\ref{eps_np}, \ref{eps_fc}, \ref{effmass_np} and \ref{effmass_fc}.

The total material energy density (as locally defined without allowance 
for gravity) will be expressible in the ``comprehensive'' treatment by
{\be \UU_{\rm mat} = \tilde\UU_{\rm kin} + \tilde\UU_{\rm int}\, , \hskip 1 cm
\tilde\UU_{\rm int} =\tilde\UU_{\rm ins} + \tilde\UU_{\rm ent} \, , \fe}
with the kinetic contribution given by the usual formula
{\be \tilde\UU_{\rm kin} = \frac{_1}{^2} \mm \big( \nn_\rmn \vv_\rmn^2 +
\nn_\rmp \vv_\rmp^2 \big)\, , \fe}
while from equations \rf{68}\fr, \rf{75}\fr and \rf{90}\fr it can be seen 
that the two fluid equation of state will be expressible in terms of 
a pair of 3-variable functions  
{\bb\tilde\UU_{\rm ins}\{ \nn_\rmI,\nn_\rmn, \nn_\rmp\}\fb} and
{\bb \aalpha\{ \nn_\rmI,\nn_\rmn, \nn_\rmp\}\fb} in terms of the
quantity that has been loosely referred by Prix \textit{et al.} 
\cite{PrixComAnd, Prix01, Rieutord02, Prix02}
to  as the ``internal energy density'' -- but that is actually the negative of 
the internal action density -- will be given by
{\be - \tilde\Lamb_{\rm int} = \tilde\UU_{\rm ins} +\aalpha\,{\Deltav}^2 
\, .\fe}
whereas the true internal energy density will be given by
 {\be \tilde\UU_{\rm int} = \tilde \UU_{\rm ins}
-\aalpha\, {\Deltav}^2   \, .\fe}
 
In the high density core region the static energy contribution 
{\bb\tilde\UU_{\rm ins}\fb} will cease to have any dependence on the ionic 
number density {\bb\nn_\rmI\fb}, not because of any tendency of 
{\bb\nn_\rmI\fb} to vanish on the lower boundary of the crust (where it 
merely ceases to be well defined) but rather because the corresponding 
clustering energy coefficient {\bb\mmu^\rmI\fb} will vanish there, so that 
the value of {\bb\nn_\rmI\fb} no longer matters.

As depicted in figure \ref{eps_np}, the entrainment coefficients
between the neutron and proton fluids in the core of a neutron star are 
expected to be very small compared to the previously neglected entrainment
coefficients in the inner crust layers. This strong entrainment in the inner
crust may have significant effects on the dynamical evolution of the star.
For instance, as recently shown by Andersson {\it et al.}
\cite{Andersson04}, sufficiently large negative values of the entrainment 
coefficients {\bb \varepsilon_\rmn \fb} and {\bb \varepsilon_\rmp \fb} 
could trigger a Kelvin-Helmholtz instability in the fluid mixture which 
might be at the origin of observed pulsar glitches. The development of 
gravitational wave detectors could provide valuable constraints upon the 
internal structure of a neutron star by asteroseismology \cite{Andersson03}. 
However a thorough understanding of the oscillation modes of a neutron star 
is required in order to extract useful information from the gravitational 
wave data. Hydrodynamical simulations of neutron star cores indicate that 
these oscillation modes are very sensitive to entrainment 
\cite{Andersson03}. The inner crust, where the two fluids are expected to 
be the most strongly coupled, should therefore be given careful
consideration in neutron star dynamical studies. As shown in the present 
work, a simple neutron star two fluid model allowing for the presence of 
the crust (but neglecting the stress anisotropy), can be easily 
implemented by suitable adjustments of the corresponding entrainment 
coefficients.

\vfill\eject

\noindent
{\bf Acknowledgements}. This work was supported by the 
LEA Astro-PF program. Pawel Haensel was supported by the KBN grant 
no. 1-P03D-008-27. Nicolas Chamel was supported
by the Lavoisier program of the French Ministry of Foreign Affairs.

\end{document}